\documentclass[conference]{IEEEtran}
\IEEEoverridecommandlockouts
\usepackage{amsmath,amssymb,amsfonts}
\usepackage{algorithmic}
\usepackage{graphicx}
\usepackage{textcomp}
\usepackage{xcolor}
\def\BibTeX{{\rm B\kern-.05em{\sc i\kern-.025em b}\kern-.08em
    T\kern-.1667em\lower.7ex\hbox{E}\kern-.125emX}}
    
\usepackage[sort, numbers]{natbib}

\usepackage{comment}

\usepackage[stable]{footmisc}

\usepackage{hyperref}
\usepackage{cleveref}[2012/02/15]

\usepackage[euler]{textgreek}
\usepackage{tipa}

\usepackage{multirow}
\usepackage{amsxtra}
\usepackage{threeparttable}
\usepackage{array}
\usepackage{longtable}

\usepackage{array}
\newcolumntype{C}[1]{>{\centering\arraybackslash}p{#1}}
\usepackage{graphicx} 
\usepackage{booktabs}

\renewcommand{\baselinestretch}{0.99} 

\def\BibTeX{{\rm B\kern-.05em{\sc i\kern-.025em b}\kern-.08em
    T\kern-.1667em\lower.7ex\hbox{E}\kern-.125emX}}
\begin{document}

\title{EMVD dataset: a dataset of extreme vocal distortion techniques used in heavy metal}

\author{\IEEEauthorblockN{Modan Tailleur}
\IEEEauthorblockA{\textit{Nantes Université, École Centrale Nantes} \\
\textit{CNRS, LS2N, UMR 6004}\\
Nantes, France \\
modan.tailleur@ls2n.fr}
\and
\IEEEauthorblockN{Julien Pinquier}
\IEEEauthorblockA{\textit{IRIT, Université de Toulouse} \\
\textit{CNRS, UT3}\\
Toulouse, France \\
julien.pinquier@irit.fr}
\and
\IEEEauthorblockN{Laurent Millot}
\IEEEauthorblockA{\textit{Institut Acte, UR 7539} \\
\textit{ENS Louis-Lumière}\\
Saint-Denis, France \\
l.millot@ens-louis-lumiere.fr}
\and
\IEEEauthorblockN{Corsin Vogel}
\IEEEauthorblockA{
\textit{ENS Louis-Lumière}\\
Saint-Denis, France \\
c.vogel@ens-louis-lumiere.fr}
\and
\IEEEauthorblockN{Mathieu Lagrange}
\IEEEauthorblockA{\textit{Nantes Université, École Centrale Nantes} \\
\textit{CNRS, LS2N, UMR 6004}\\
F-44000 Nantes, France \\
mathieu.lagrange@ls2n.fr}
}

\maketitle

\begin{abstract}
In this paper, we introduce the Extreme Metal Vocals Dataset, which comprises a collection of recordings of extreme vocal techniques performed within the realm of heavy metal music. The dataset consists of 760 audio excerpts of 1 second to 30 seconds long, totaling about 100 min of audio material, roughly composed of 60 minutes of distorted voices and 40 minutes of clear voice recordings. These vocal recordings are from 27 different singers and are provided without accompanying musical instruments or post-processing effects. The distortion taxonomy within this dataset encompasses four distinct distortion techniques and three vocal effects, all performed in different pitch ranges. Performance of a state-of-the-art deep learning model is evaluated for two different classification tasks related to vocal techniques, demonstrating the potential of this resource for the audio processing community.
\end{abstract}

\section{Introduction}
\label{sec:Intro}
Heavy metal music, known in part for its distinctive vocal styles, often features singers who employ vocal distortion techniques to imbue their performances with unique timbres and heightened aggression. These techniques manifest as unconventional vibratory behaviors within the vocal tract and other larynx components such as the ventricular folds. The complexity of these techniques gives rise to a vast diversity of vocal timbres and production methods. Notably, the most extreme subgenres of heavy metal, such as death metal, black metal, metalcore, and grindcore, exhibit quite distinct vocal approaches. 

Over time, a variety of extreme vocals taxonomies have emerged. Nieto \cite{nieto_voice_2008}\cite{nieto_unsupervised_2013} defined a taxonomy of the most extreme vocal techniques into three categories: the death growl (predominantly used in the death metal subgenre), the fry scream exhale (commonly associated with the metalcore subgenre), and the fry scream inhale (typically employed in grindcore and deathcore). Similarly, Sadolin \cite{sadolin_complete_2000} and Thuesen et al. \cite{thuesen_curbingmetallic_2017} introduced related albeit less comprehensive taxonomies of extreme vocalizations prevalent in heavy metal music. Recent research by Hainaut \cite{hainaut_vocalites_2020} unveiled significant distinctions between vocal styles employed in death metal and those in black metal, despite their apparent reliance on the same overarching technique. Furthermore, Purcell \cite{purcell_death_2003} describes an additional facet of metal vocalization, denoted as "sepulchral," characterized by even deeper tonalities, often rendering lyrics unintelligible. A more anatomical approach was taken by Chevaillier et al. \cite{chevaillier_voix_2011} who categorized metal voices into four distinct classes, each correlated with specific anatomical regions of the oral cavity and vocal tract anatomy. Those studies will root the proposal of the taxonomy considered for this dataset, as detailed in section \ref{sec:ExVocTax}.

Despite the emergence of several sound detection studies closely related to extreme heavy metal vocalization over the last two decades \cite{huang_scream_2010,ishi_method_2007,hayasaka_noise-robust_2017,pohjalainen_shout_2011,nandwana_robust_2015,laffitte_deep_2016}, and the numerous acoustic analysis of growls and screams in heavy metal~\cite{smialek_spectrographic_2012,guzman_noriega_aerodynamic_2018,kato_acoustic_2013,eckers_voice_2009,smialek_musical_2012}, the translation of this research into practical applications within the music realm remains a challenge. We believe that the study of \textit{heavy metal} vocalization is promising specifically for vocal technique classification, which finds application in automatic content labeling for streaming platforms \cite{nieto_unsupervised_2013} or to enhance live performances by applying vocal effects in real-time. Furthermore, it offers the potential to advance voice distortion generative algorithms \cite{gentilucci_vocal_2018}. The emergence of deep learning algorithms presents an opportunity to elevate the capabilities of existing methodologies. 

Unfortunately, the development of these applications encounters a significant obstacle: the scarcity of extensive and robust datasets. To address this gap, we present the Extreme Metal Vocals Dataset (EMVD)\footnote{Dataset available at: \url{https://zenodo.org/record/8406322}, in line with the open science policy of the European Union.}, a comprehensive and diverse collection of extreme vocal performances. This dataset features a new taxonomy that encompasses four distinct distortion techniques across three vocal ranges, along with three vocal effects,  making it broader in scope compared to existing datasets~\cite{kalbag_scream_2022}. Unlike the dataset proposed by Kalbag and Lerch \cite{kalbag_scream_2022}, the EMVD dataset comprises unprocessed vocal recordings devoid of musical accompaniment. It includes contributions from 27 different vocalists, representing a substantial improvement over the Soundiron dataset \cite{soundiron_voices_nodate}, which comprises vocals from 5 singers. As of now, it stands as the largest and most varied dataset of its kind, tailored specifically for analyzing vocal techniques within the most extreme heavy metal subgenres. Section \ref{sec:ExVocTax} introduces our novel taxonomy for extreme vocal techniques. Moving forward to section \ref{sec:ExVocDa}, we provide a comprehensive overview of the dataset creation process, encompassing recording procedures and metadata collection. 
In order to illustrate the potential of our dataset, we present the results of a basic model\footnote{Code available at: \url{https://github.com/modantailleur/ExtremeMetalVocalsDataset}} adapted to the classification of techniques in section \ref{sec:BaSy}.

\section{Taxonomy}
\label{sec:ExVocTax}

This dataset primarily focuses on clear vocals (i.e., unsaturated vocals) and on the most extreme vocal distortion techniques that have been studied, excluding lighter distortion techniques such as the rattle, the growl, or the distortion. For a comprehensive understanding of these lighter techniques, please refer to Sadolin's taxonomy \cite{sadolin_complete_2000}. As detailed in Section~\ref{sec:Intro}, the categorization of extreme vocal techniques within the realm of \textit{heavy metal} music has historically been empirical, resulting in various taxonomies among different researchers. Taking the most important aspects of those proposals, this study introduces yet another taxonomy that aims at encompassing the various techniques while emphasizing their association with specific extreme subgenres within heavy metal. This taxonomy also breaks down into different vocal registers for each technique: high register for a voice approaching the limit of the head voice, mid register for the singer's most comfortable tonal range, and low register for a significantly lower-pitched voice\footnote{Audio examples available at: \url{https://modantailleur.github.io/ExtremeMetalVocalsDataset/}}. See table \ref{tab:taxo} for a quick overview of the taxonomy.

\subsection{Vocal Techniques}
\label{sec:ExVocTax-VocTe}
The following techniques serve as alternatives to the clear voice in heavy metal music, in the sense that they're offering the ability to convey intelligible lyrics and accommodate a range of tonal variations.

\subsubsection{Black Shriek}
The Black Shriek finds widespread utilization in black metal and death metal. It is typically characterized by a higher pitch compared to the Death Growl, particularly noticeable in vowel pronunciation, producing a timbre that suggests an intermediary vowel sound between [a] (as in "cat") and [\textepsilon] (as in "mother")~\cite{hainaut_vocalites_2020}. This vocal technique is notably employed by vocalists in bands such as \textit{Marduk}, \textit{Satyricon}, \textit{Behemoth}, \textit{Dimmu Borgir}, and \textit{Darkthrone}.

\subsubsection{Death Growl}
The Death Growl is predominantly employed in death metal but also appears in thrash metal and black metal vocalisations. This technique is distinguished by an intermediate vowel sound situated between [a] and [\textopeno] (as in "all")~\cite{hainaut_vocalites_2020}. Vocalists in groups such as \textit{Cannibal Corpse}, \textit{Cattle Decapitation}, \textit{Deicide}, and \textit{Gorod} are known for employing the death growl.

\subsubsection{Hardcore Scream}
The Hardcore Scream, associated with hardcore and grindcore, entails a vociferous shout characterized by varying degrees of voice distortion. This technique is also employed in the high vocal register in numerous black metal compositions. Noteworthy vocalists, including those in bands such as \textit{Slipknot}, \textit{The Dillinger Escape Plan}, \textit{Stick To Your Guns}, \textit{Fall In Archaea}, \textit{Terror}, \textit{Get The Shot}, and \textit{Silverstein}, have employed the hardcore scream.

\subsubsection{Grind Inhale}
The \textit{Grind Inhale} entails a screaming technique produced by inhaling air. While it finds application across various metal genres, it is predominantly used in grindcore, although it remains a relatively secondary technique compared to those mentioned above. This vocal technique is exemplified by vocalists in bands such as \textit{Annotations Of An Autopsy}, \textit{Walking The Cadaver}, and \textit{Archspire}.

\subsection{Vocal Effects}
Heavy metal vocalists also incorporate vocal effects to punctuate their performances. These effects are often only sparingly employed within musical compositions, as they render speech entirely unintelligible, setting them apart from the vocal techniques presented in the previous section.

\subsubsection{Pig Squeal}
The Pig Squeal technique involves imitating a pig's call and is typically characterized by high-pitched tones. Similar to the Grind Inhale, it is primarily produced by inhaling air rather than exhaling.

\subsubsection{Deep Gutturals}
The Deep Gutturals represent Death Growls executed in the extreme lower tessitura of the singer's range, aiming to achieve the lowest possible vocalization. 

\subsubsection{Tunnel Throat}
The Tunnel Throat technique involves curling the tongue against the palate to produce a profoundly low-pitched growl. The resulting sound is often likened to the noise of an unblocking sink, with a pitch slightly higher than that of Deep Gutturals.


\begin{table}[!htb]
\caption{Taxonomy of Vocal Techniques and Vocal Effects}
\label{tab:taxo}
\begin{tabular}{C{0.3cm}C{0.7cm}|C{1.1cm}C{1.2cm}C{2cm}|}
\cline{3-5}
& & Breathing method & Registers & Main metal subgenres \\ \cline{2-5}
\multirow{4}{*}{\raisebox{-1.7\height}{\rotatebox[origin=c]{90}{\textbf{techniques}}}} & \multicolumn{1}{|l|}{Black Shriek} & Exhale & High, Mid & black metal, death metal \\ \cline{2-5}
 & \multicolumn{1}{|l|}{Death Growl} & Exhale & Mid, Low & death metal, thrash metal, black metal \\ \cline{2-5}
 & \multicolumn{1}{|l|}{Hardcore Scream} & Exhale & High, Mid, Low & hardcore, grindcore \\ \cline{2-5}
 & \multicolumn{1}{|l|}{Grind Inhale} & Inhale & High, Mid, Low & grindcore \\ \cline{2-5} \\  \cline{2-5}
\multirow{3}{*}{\raisebox{-0.5\height}{\rotatebox[origin=c]{90}{\textbf{effects}}}} & \multicolumn{1}{|l|}{Pig Squeal} & Inhale & - & - \\ \cline{2-5}
 & \multicolumn{1}{|l|}{Deep Gutturals} & Exhale & - & - \\ \cline{2-5}
 & \multicolumn{1}{|l|}{Tunnel Throat} & Exhale & - & - \\ \cline{2-5}
\end{tabular}
\end{table}

\section{The EMVD dataset}
\label{sec:ExVocDa}

\begin{figure*}[!htb]
\begin{center}
\includegraphics[width=\linewidth]{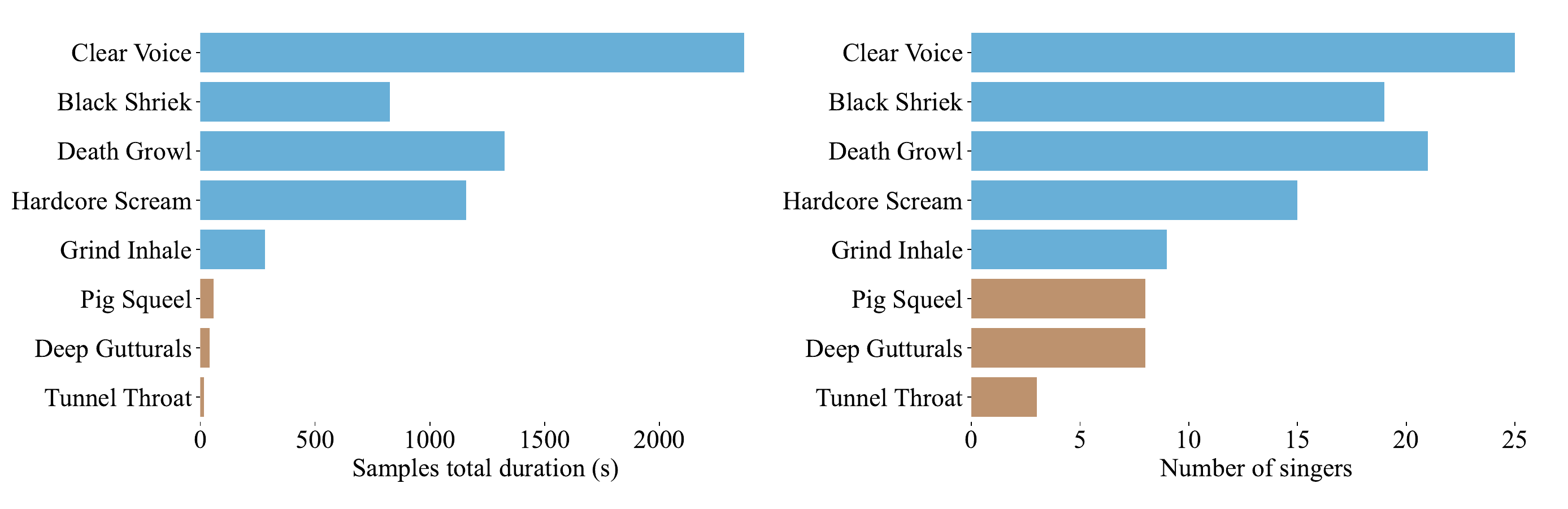}
\end{center}
\vspace{-0.5cm}
\caption{EMVD corpus details}
\vspace{-0.4cm}
\label{fig:technique_duration}
\end{figure*}

\subsection{Recording devices}
\label{subsec:Equip}
For the recording sessions, a mobile setup was selected to accommodate as many singers as possible. An SM58 microphone was employed, chosen for its prevalence as a microphone commonly used by metal singers during live performances. A closed-back headphone served for music playback and provided the singers with a monitor of their own voice if they desired to hear it during recording. An audio interface \textit{Scarlett 6i6} by \textit{Focusrite} was responsible for connecting the laptop, microphone, and headset.

In some cases, singers were recorded remotely using their own equipment (a stage microphone and an audio interface) which are documented in the database. These singers were provided with a video tutorial and explanatory documents to facilitate their participation in the project, with the main author remotely guiding them.

\subsection{Recording procedure}

Each singer was instructed to sustain three vowels—[a] as in "cat," [i] as in "ship," and [u] as in "book"—for a duration of five seconds each. They were required to maintain a consistent pitch not only within each vowel but also across all vowels produced. After this, they were asked to perform for approximately 15 seconds using the same vocal technique, but this time with lyrics of their choosing. The lyrics had to remain the same across all technique categories. A musical loop was provided in the singers' headphones during each recording. Each vocal technique was recorded across several registers (high, mid, and low) depending on their relevance to the specific technique. If a singer was uncomfortable using a particular technique or effect, they were explicitly asked not to use it to prevent potential vocal strain. Notably, the Grind Inhale technique, although producible in multiple registers, was recorded in only one register, as many singers deemed it potentially harmful to their voice.


Of the 27 singers involved in the project, 22 were recorded using the equipment detailed in section \ref{subsec:Equip}, while 5 singers utilized their own remote recording setups.

\subsection{Metadata}


Based on his experience with this type of content, the main author provided grades to individual audio files created by the singers, ranging from 0 to 2. A 2 grade suggests that the technique closely represents the intended vocal technique, 1 indicates that it moderately represents the vocal technique, and 0 signifies that the technique does not adequately represent the vocal technique. Audio files rated as 0 should not be employed for deep learning applications, but they are retained within the dataset in case future re-evaluation of the audio files is desired. Notably, approximately 70\% of the dataset's audio files received grades of 2 or 1 from the authors and are thus suitable for being used in diverse applications.

\begin{figure*}[!htb]
\begin{center}
\includegraphics[width=\linewidth]{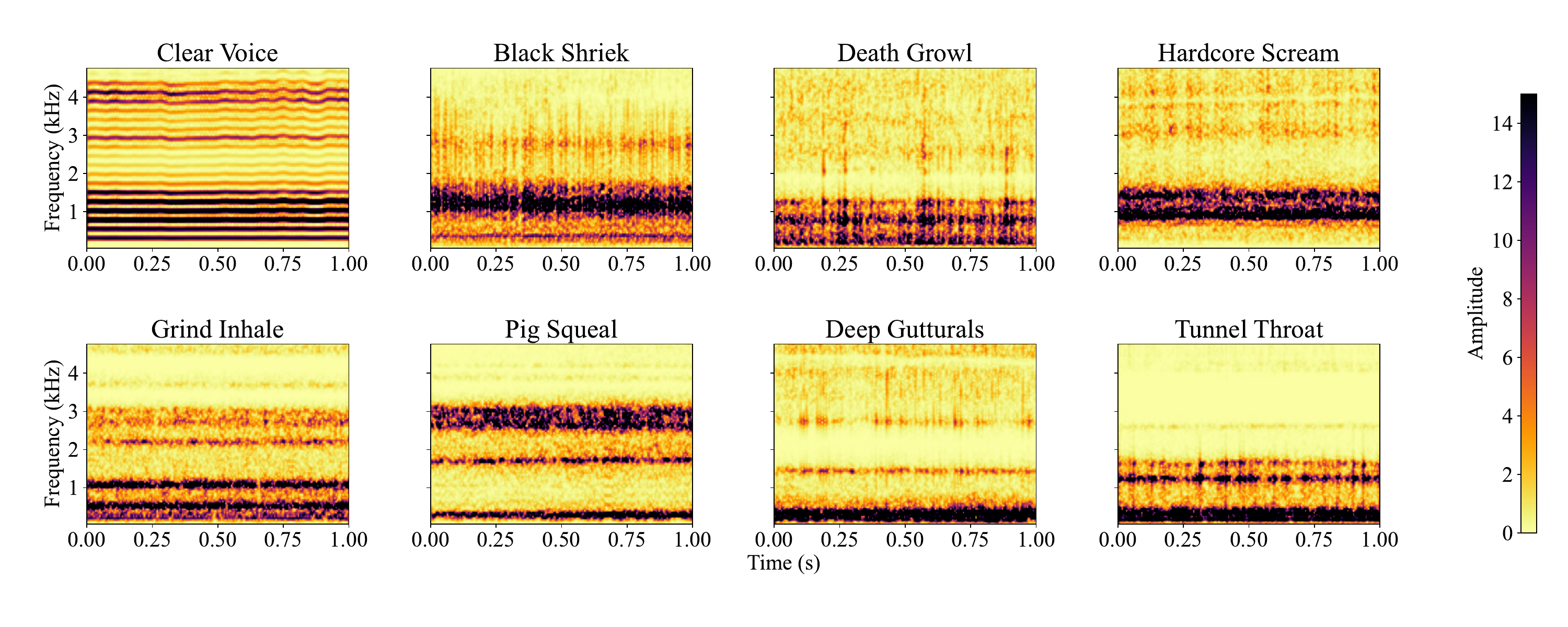}
\end{center}
\vspace{-1.1cm}
\caption{Example of a 1s log-Mel-spectrogram for each vocal technique and effect}
\vspace{-0.4cm}
\label{fig:spectrograms_fb}
\end{figure*}

\subsection{Statistics and examples}

Keeping only 2 and 1 graded files with the main author grading system, the dataset comprises a total of 760 individual audio files, which corresponds to a duration of 100 minutes of audio recordings. Figure \ref{fig:technique_duration} provides a breakdown of the recording times for each vocal technique and vocal effect. The dataset thus consists of 38\% Clear Voice recordings, 13\% Black Shriek, 21\% Death Growl, 19\% Hardcore Scream, and 5\% Grind Inhale. Vocal effects constitute less than 2\% of the overall dataset. Figure \ref{fig:technique_duration} also illustrates the number of distinct singers who participated in the recordings for each vocal technique and vocal effect. Notably, the techniques of Clear Voice, Black Shriek, Death Growl, and Hardcore Scream involved recordings by more than 15 different singers each. In contrast, the Grind Inhale technique and vocal effects were recorded by fewer than 10 singers each. Among the 27 singers that participated to the study, it is worth to note that only 4 were women.

Figure \ref{fig:spectrograms_fb} presents an example of audio spectrogram for each of the 5 vocal techniques and 3 vocal effects. These figures reveal that distinguishing Clear Voice from vocal distortion techniques is expected to be much easier than differentiating between various distortion techniques, as the harmonics of Clear Voice are completely altered in the other techniques. 

A noteworthy comparison can be made between Grind Inhale and Pig Squeal techniques. Despite both involving inhaling air, Pig Squeal exhibits a significantly higher level of noise around 2.8 kHz. This noise contributes to its resemblance to the sound of a squealing pig. 

Deep Gutturals and Tunnel Throat exhibit very similar spectrograms, with the primary distinction being the elevated energy in frequencies above 900 Hz, which is more pronounced in Tunnel Throat. Additionally, these frequencies appear to be modulated at a certain rate, contributing significantly to the generation of a sound reminiscent of a sink being unblocked. 

Furthermore, Black Shriek appears to encompass characteristics from both Death Growl and Hardcore Scream techniques, featuring localized noise typical of Hardcore Scream as well as lower, more distorted noise akin to Death Growl.

\section{Vocal Techniques Classification}
\label{sec:BaSy}

Due to limited data availability for vocal effects and the Grind Inhale technique, only Clear Voice, Hardcore Scream, Black Shriek, and Death Growl are retained for experimentation. 

Two classification tasks are considered. First, we train an EfficientNet  for binary classification, focusing on distinguishing Clear Voice from distorted vocals. Second, we train another EfficientNet for multi-class classification, aiming to distinguish among the four classes (Clear Voice, Black Shriek, Death Growl, Hardcore Scream). 

An EfficientNet $b_0$ \cite{koonce_efficientnet_2021} architecture is trained on Mel spectrograms derived from 1-second audio samples sampled at 48 kHz. The spectrograms are created using a frame length of 1024 samples, a hop length of 256, 128 Mel-frequency bins, a minimum frequency of 20 Hz, and a maximum frequency of 8~kHz. Only audio files ranked with a 2 or a 1 by the authors are included. A k-fold cross-validation with 4 folds is performed, reserving 20\% of the training data of each fold for validation. All audio files are normalized before processing. The model is trained with a learning rate of $10^{-3}$ with adam's optimizer using the BCE (Binary Cross-Entropy) loss for 20 epochs and a batch size of 32, which corresponds to 1680 iterations. The model associated with the best checkpoint is retained for further evaluation. Subsequently, we compute the accuracy scores of this optimal model's performance on the evaluation dataset for each cross-validation split.

\begin{table}[!htb]
\caption{Accuracy scores on classification tasks}
\begin{tabular}{c|l|l|}
\cline{2-3}
\multicolumn{1}{l|}{}                                                                                   & Micro Accuracy & Macro Accuracy \\ \hline

\multicolumn{1}{|c|}{\begin{tabular}[c]{@{}c@{}}Binary Classification \end{tabular}} & 93\%          & 93\%          \\ \hline
\multicolumn{1}{|c|}{Multi-class classification}                                                          & 75\%         & 70\%         \\ \hline
\end{tabular}
\label{tab:classif_results}
\end{table}

\begin{figure}[!htb]
\begin{center}
\includegraphics[width=\linewidth]{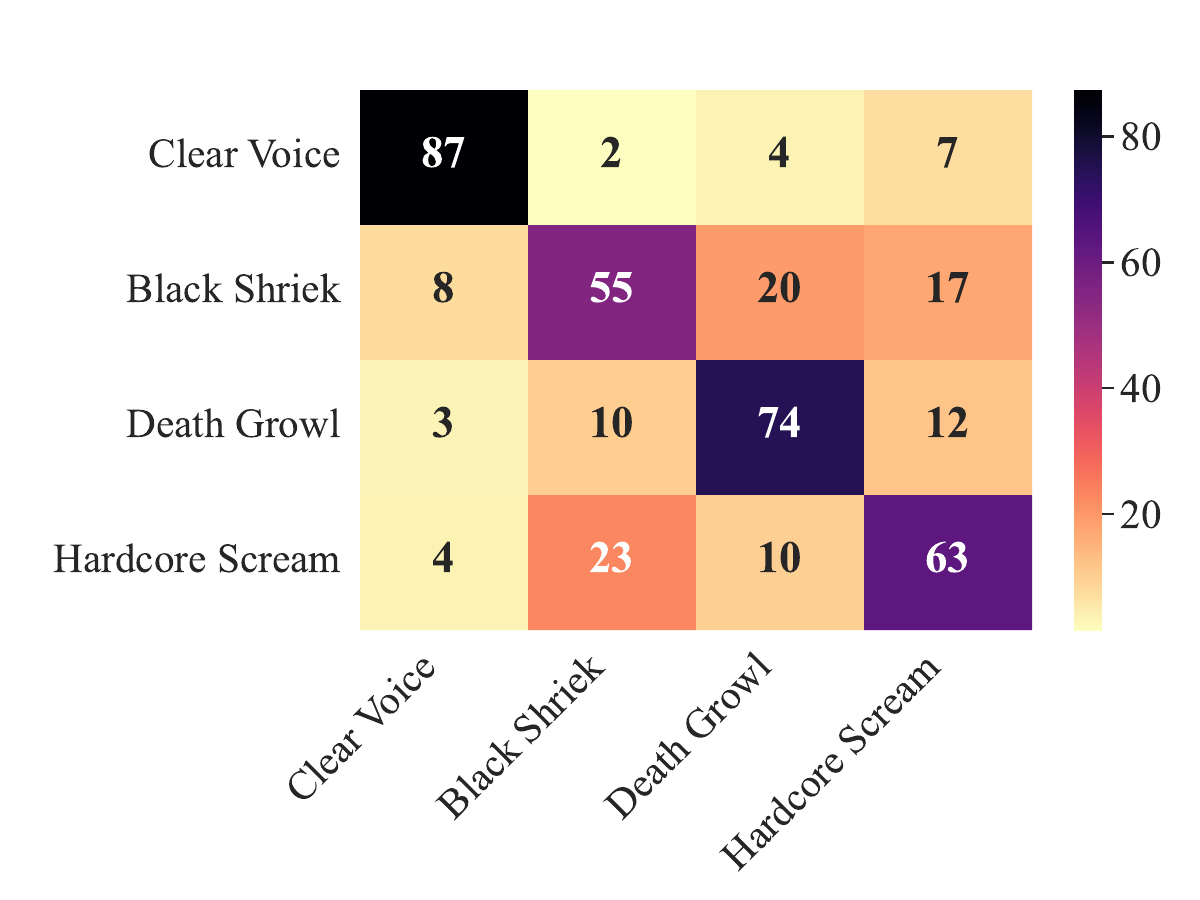}
\end{center}
\vspace{-0.5cm}
\caption{Confusion matrix for multi-class classification (values rounded to nearest unit).}
\vspace{-0.1cm}
\label{fig:mat_conf}
\end{figure}

Table \ref{tab:classif_results} summarizes the performance of the model for both multi-class and binary classification, indicating typical micro and macro accuracy scores. The confusion matrix in Figure \ref{fig:mat_conf} further illustrates the classification results for the multi-class classification. The binary classification task achieve a notable Micro Accuracy score of 93\% and a Macro Accuracy score of 93\%. Conversely, the multi-class classification achieve a Micro Accuracy score of 75\% and a Macro Accuracy score of 70\%. This discrepancy can be attributed to the inherent spectrogram characteristics, with Clear Voice being significantly easier to distinguish from the other techniques. Notably, the confusion matrix shows that the Black Shriek appears challenging to differentiate, possibly because it exhibits characteristics of both Death Growl and Hardcore Scream. Those characteristics highly depend on the singer, as some singers produced Black Shriek techniques that resemble more the Death Growl, and others that ressemble more the Hardcore Scream. In a future work, the Black Shriek audio samples could be reannotated by expert singers to fit into the Death Growl or the Hardcore Scream categories.


\section{Conclusion}
\label{sec:Concl}
We hope that our EMVD datatset will represent a significant milestone in the domain of extreme metal vocalization datasets. This extensive collection, totaling 100 minutes of raw, background-free audio, stands as the largest and most diverse dataset of its kind to date since it comprises 740 audio files from 27 unique vocalists, encompassing five distinct vocal techniques and three vocal effects.


The potential applications of this dataset are manifold. First, it can be considered as a training resource to aid aspiring vocalists in honing their skills by providing a unique tool for technique evaluation and improvement. We also anticipate that this dataset will help further the understanding and utilization of extreme metal vocalization.

In terms of audio processing research, we believe it can be a valuable resource for pioneering new avenues of investigation, including technique recognition for automated tagging, and real-time vocal effect processing during live performances based on online detection of the vocalization type.

\section*{Acknowledgments}
We want to thank Oriol Nieto, Geoffroy Peeters, Christophe d'Alessandro and Boris Doval for fruitful discussion. We particularly want to thank Joshua Smith for guidance for the design of the taxonomy. 

\bibliographystyle{IEEEtran}
\bibliography{Bibliographie}



\end{document}